# On the Role of Side Information In Strategic Communication

Emrah Akyol, Cedric Langbort, and Tamer Başar
{akyol, langbort, basar1}@illinois.edu
University of Illinois at Urbana-Champaign

*Abstract*—This paper analyzes the fundamental limits of strategic communication in network settings. Strategic communication differs from the conventional communication paradigms in information theory since it involves different objectives for the encoder and the decoder, which are aware of this mismatch and act accordingly. This leads to a Stackelberg game where both agents commit to their mappings ex-ante. Building on our prior work on the point-to-point setting, this paper studies the compression and communication problems with the receiver and/or transmitter side information setting. The equilibrium strategies and associated costs are characterized for the Gaussian variables with quadratic cost functions. Several questions on the benefit of side information in source and joint source-channel coding in such strategic settings are analyzed. Our analysis has uncovered an interesting result on optimality of uncoded mappings in strategic source-channel coding in networks.

## I. INTRODUCTION

Consider a Stackelberg game between a transmitter and a receiver which have conflicting objectives. Both agents commit to their mappings ex-ante, i.e., they decide the mappings based on the source/channel statistics. Such communication settings where the objectives of the transmitter and the receiver differ are referred to here as "strategic communication." While such settings have been studied extensively in the economics literature (see e.g., [1], [2]), an information theoretic analysis of strategic communication has been undertaken only very recently [3].

This work extends the strategic compression and joint source-channel coding problems in point-to-point settings (which were analyzed in our prior work [3]) to ones with receiver/transmitter side information (SI). We consider the strategic equivalents of the well-known Wyner and Ziv [4] setting, and also study the associated source-channel coding settings. Strategic aspect of the problem yields some interesting variations of the well-known results associated with non-strategic settings. For example, the celebrated result of Wyner and Ziv on the absence of rate loss in quadratic-Gaussian setting [4] (due to the non-availability of SI at the transmitter side) also holds in strategic settings, i.e., the presence of decoder SI at the encoder cannot be helpful (to the transmitter) in strategic Wyner-Ziv. Another surprising result, due to Goblick [5], is that single-letter linear coding is optimal for joint source-channel coding of a Gaussian source over an additive Gaussian channel for quadratic costs (distortion and channel cost). It is well-understood that this optimality breaks down when there is receiver SI. We analyze the strategic equivalent of these settings and show that there exist problem parameters such that single-letter linear strategies continue to possess optimality even in the receiver SI settings.

We note in passing that the question of source compression with mismatched distortion measures has been addressed before, see e.g., [6], and the references therein. The main difference between our line of "strategic communication" work and all earlier ones in information theory is that in our problem, the encoder and the decoder are aware of the mismatched objectives, and they act (design the encoding and the decoding mappings) accordingly. In prior work, this mismatch was considered to be created by nature (worst case, or robust design) [6], [7] or by an adversarial secondary decoder [8], but not as an intentional consequence of strategic agents.

## II. PRELIMINARIES

### A. Notation

$\mathbb{R}$ and $\mathbb{R}^+$ denote the respective sets of real numbers and positive real numbers. Let $\mathbb{E}(\cdot)$ denote the expectation operator. The Gaussian density with mean $\mu$ and variance $\sigma^2$ is denoted by $\mathcal{N}(\mu, \sigma^2)$. All logarithms in the paper are natural logarithms and may in general be complex valued, and the integrals are, in general, Lebesgue integrals. $\mathcal{S}$ denotes the set of Borel measurable, square integrable functions $\{f : \mathbb{R} \to \mathbb{R}\}$. We use standard information theoretic and game theoretic notations for the related quantities throughout this paper (cf. [9], [10]).

### B. An Overview of Point-to-Point Results

Consider the general communication system whose block diagram is shown in Figure 1. The source $X$ and private information $\theta$ are mapped into $U \in \mathbb{R}$ which is fully determined by the conditional distribution $p(\cdot|x, \theta)$. For the sake of brevity, and with a slight abuse of notation, we refer to this as a stochastic mapping $U = g(X, \theta)$ so that

$$\mathbb{P}(g(X, \theta) \in \mathcal{U}) = \int_{u' \in \mathcal{U}} p(u'|x, \theta) \mathrm{d}x \mathrm{d}\theta \quad \forall \mathcal{U} \subseteq \mathbb{R} \quad (1)$$

holds almost everywhere in $X$ and $\theta$. Let the set of all such mappings be denoted by $\Gamma$ (which has a one-to-one correspondence to the set of all the conditional distributions that construct the transmitter output $U$).

We consider an additive noise channel as shown in Figure 1, with Gaussian noise $N \sim \mathcal{N}(0, \sigma_N^2)$, hence the input to the receiver is $Y = U + N$. We first consider the simpler problem where there is no channel noise, i.e., we effectively assume $\sigma_N^2 = 0$, and hence $Y = U$ (almost everywhere). The receiver produces an estimate of the source $\hat{X}$ through a mapping $h \in \mathcal{S}$ as $\hat{X} = h(Y)$. The objective of the receiver is to minimize

$$D_D = \mathbb{E}\{d_D(X, \hat{X})\} \qquad (2)$$

while that of the transmitter is to minimize

$$D_E = \mathbb{E}\{d_E(X, \theta, \hat{X})\} \qquad (3)$$

over the mappings $g(\cdot,\cdot) \in \Gamma, h(\cdot) \in \mathcal{S}$. In game theoretic terms, we consider a *Stackelberg game*, where the transmitter (the leader) knows that the decoder (the follower) acts to minimize its own measure in (2) as a function of encoding mapping $g(\cdot,\cdot)$. From the transmitter's point of view, we are looking for an encoding mapping, $g(\cdot,\cdot)$, that minimizes a distortion measured by $d_E$, with a decoder $h(\cdot)$ matched to the distortion measure $d_D$.

**Quadratic-Gaussian Setting**: Most of our results concern the setting where the source and the private information are jointly Gaussian i.e., $(X, \theta) \sim \mathcal{N}(0, R_{X\theta})$ where, without any loss of generality, $R_{X\theta}$ is parametrized as $R_{X\theta} = \sigma_X^2 \begin{bmatrix} 1 & \rho \\ \rho & r \end{bmatrix}$, with $r > \rho^2$, and the distortion measures are given as follows:

$$d_E(x, \theta, y) = (x + \theta - y)^2; \quad d_D(x, y) = (x - y)^2. \qquad (4)$$

Hence, we have the following cost functions:

$$D_E = \mathbb{E}\{(X + \theta - \hat{X})^2\}; \quad D_D = \mathbb{E}\{(X - \hat{X})^2\}. \qquad (5)$$

The following theorem characterizes the equilibrium in the noiseless quadratic-Gaussian (Q-G) setting

**Theorem 1** ( [3]). *In the noiseless Q-G setting, the unique equilibrium is achieved by $g(X, \theta) = X + \alpha\theta$ and $h(Y) = \kappa Y$ where $\alpha$ and $\kappa$ are constants given as:*

$$\alpha = \frac{A - 1}{2(r + \rho)}, \quad \kappa = \frac{1 + \alpha\rho}{1 + \alpha^2 r + 2\alpha\rho} \qquad (6)$$

*Distortion costs at the equilibrium are*

$$D_E = \sigma_X^2 \left(1 + \frac{(A - 3)(r + \rho)}{A - 1}\right) \qquad (7)$$

$$D_D = \sigma_X^2 \left(\frac{(r - \rho^2)(A - 1)}{A(2r + A\rho + \rho)}\right) \qquad (8)$$

*where $A = \sqrt{1 + 4(r + \rho)}$.*

**Strategic Compression**: A memoryless source $X^n$ and the private information sequence $\theta^n$ are mapped to an index set $\mathcal{M}$ by $f_E : \mathcal{X}^n \times \theta^n \longrightarrow \mathcal{M}$. The decoder applies $f_D : \mathcal{M} \longrightarrow \mathcal{Y}^n$ to generate the reconstruction sequence $\hat{X}^n$. An achievable triple $(R, D_E, D_D)$ satisfies

$$\frac{1}{n}\log|\mathcal{M}| \leq R + \delta$$
$$E\{d_E^n(X^n, \theta^n, f_D(f_E(X^n, \theta^n)))\} \leq D_E + \delta$$
$$E\{d_D^n(X^n, f_D(f_E(X^n, \theta^n)))\} \leq D_D + \delta,$$

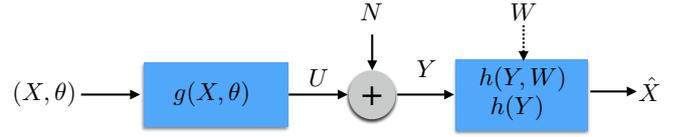

Fig. 1: The strategic variant of Gaussian test channel, with or without receiver side information $W$.

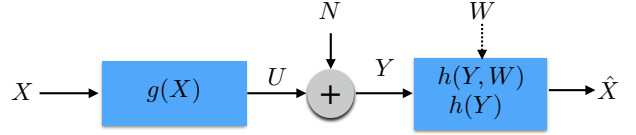

Fig. 2: The (non-strategic) Gaussian test channel, with or without receiver side information $W$.

for every $\delta > 0$ and sufficiently large $n$. The set of achievable R-D triples $(R, D_E, D_D)$ is denoted here as $\mathcal{RD}_S$ which is characterized in the following theorem.

**Theorem 2** ( [3]). *$\mathcal{RD}_S$ is the convex hull of the set of all triplets $(R, D_E, D_D)$ for which there exist a function $h : \mathcal{Y} \to \hat{\mathcal{X}}$ and a conditional distribution $p(Y|X, \theta)$ such that*

$$R \geq I(X, \theta; Y)$$

$$D_E \geq \mathbb{E}\{d_E(X, \theta, h(Y))\}, \quad D_D \geq \mathbb{E}\{d_D(X, h(Y))\}.$$

The region of $R, D_E, D_D$ at equilibrium follows from optimizing $\mathcal{R}_S$ over $p(Y|X, \theta)$ which satisfies $E\{d_E(X, \theta, h(Y))\} \leq D_E$ and $h(\cdot)$ which satisfies $\mathbb{E}\{d_D(X, h(Y))\} \leq D_D$:

$$R = \inf_{p(Y|X,\theta)} \inf_h I(X, \theta; Y) \qquad (9)$$

[ [3]] The following theorem characterizes the strategic R-D function for the quadratic-Gaussian equilibrium.

**Theorem 3.** *For the quadratic-Gaussian setting, the equilibrium $(D_E, D_D)$ pair in terms of $R$ is:*

$$D_D = \sigma_X^2 2^{-2R}\left(1 + (2^{-2R} - 1)\left(\frac{(r - \rho^2)(A - 1)}{A(2r + A\rho + \rho)}\right)\right) \qquad (10)$$

$$D_E = \sigma_X^2 \left(1 + 2\rho + r - (1 - 2^{-2R})\frac{A(r + \rho) + \rho}{A - 1}\right) \qquad (11)$$

*where $A = \sqrt{1 + 4(r + \rho)}$. The forward test channel that achieves R-D function is*

$$Y = X + \alpha\theta + S \qquad (12)$$

*where $S \sim \mathcal{N}(0, \sigma_S^2)$ is independent of $X$ and $\theta$ and $\alpha$ is given in Theorem 1.*

**Remark 1.** *Theorem 3 admits an intuitive interpretation: in the Q-G setting, strategic compression simplifies to compressing $X + \alpha\theta$ where $\alpha$ is the coefficient in the simple equilibrium. This enables, in practice, the use of standard, "of the shelf"*

*encoding codes for strategic compression operating on the effective source $X + \alpha\theta$.*

**Noisy Equilibrium:** We first review the non-strategic equivalent of the noisy equilibrium setting. Consider the general communication system whose block diagram is shown in Figure 2, the source $X \sim \mathcal{N}(0, \sigma_X^2)$ is to be transmitted to the receiver via $g \in \mathcal{S}$ as $U = g(X)$ over an additive Gaussian channel; hence the input to the receiver is $Y = U + N$, where $N \sim N(0, \sigma_N^2)$ is statistically independent of $X$. The receiver produces its output $\hat{X}$ through an $h \in \mathcal{S}$ as $\hat{X} = h(Y)$. The common objective of both agents is to minimize $\mathbb{E}(X - \hat{X})^2$, while the transmitter has an average power constraint $\mathbb{E}\{U_i^2\} \leq P_T$. The following result due to Goblick [5], states Shannon sense[1] optimality of single-letter linear strategies.

**Theorem 4** ( [5]). *For the Gaussian test channel problem, single-letter mappings*

$$g(X) = \sqrt{\frac{P_T}{\sigma_X^2}}X, \quad h(Y) = \frac{\sigma_X^2}{P_T + \sigma_N^2}\sqrt{\frac{P_T}{\sigma_X^2}}Y$$

*are the essentially unique[2], Shannon sense optimal encoding/decoding mappings.*

This optimality breaks down in the presence of receiver side information, shown as $W$ in Figure 2, and linear strategies are no longer optimal even in the zero-delay case (see e.g., [11]).

The following theorem states a similar optimality result in the strategic version of the problem (without SI).

**Theorem 5** ( [3]). *For the noisy Q-G equilibrium, the strategies*

$$g(X,\theta) = \sqrt{\frac{P_T}{\sigma_X^2(1+2\alpha\rho+\alpha^2 r)}}(X+\alpha\theta), \; h(Y) = \mathbb{E}\{X|Y\} \quad (13)$$

*with $\alpha = \frac{-1+\sqrt{1+4(r+\rho)}}{2(r+\rho)}$ are Shannon sense optimal for all power levels.*

In Section V, we incorporate SI into the problem setting. We show that, unlike its non-strategic counterpart which does not admit a linear optimal solution, there exist values for the problem parameters that render the solution to be linear.

### III. NOISELESS EQUILIBRIUM

We begin with noiseless equilibrium with receiver SI. The focus of our results is the quadratic-Gaussian setting, i.e., $(X, \theta, W) \sim \mathcal{N}(0, R_{X\theta W})$ where, $R_{X\theta W}$ is parametrized as

$$R_{X\theta W} = \sigma_X^2 \begin{bmatrix} 1 & \rho_{X,\theta} & \rho_{X,W} \\ \rho_{X,\theta} & r_\theta & \rho_{\theta W} \\ \rho_{X,W} & \rho_{\theta W} & r_W \end{bmatrix}$$

---

[1]Shannon sense optimality refers to optimality within the strategies that allow asymptotically high delay. If a single-letter strategy is Shannon sense optimal, it is also optimal among all single-letter strategies, but the converse does not hold.

[2]If $g(X) = cX$ and $h(Y) = dY$ pair is a solution to this problem, $g(X) = -cX$ and $h(Y) = -dY$ is also a solution due to symmetry, which is why the solution is "essentially" unique.

and the distortion measures are given as in (4). Hence, we have (5) as the cost functions.

The following lemma states that mappings at the equilibrium are linear (affine if variables have non-zero mean).

**Lemma 1.** *The noiseless Q-G equilibrium is achieved by mappings*

$$g(X, \theta) = X + \alpha_{SI}\theta, \quad h(Y, W) = bY + cW, \quad (14)$$

*for some $\alpha_{SI}, b, c \in \mathbb{R}$.*

The proof follows identical steps to those of Theorem 1, and hence is omitted here. The coefficients, $\alpha_{SI}, b, c$ at this equilibrium can be explicitly computed as in the case of Theorem 1, but this computation is rather involved and not included here. Instead, we focus on the high level impact of SI. We introduce $D_E^{SI}$ and $D_D^{SI}$ denote the distortion of the transmitter and the receiver at the equilibrium.

Next, we analyze the benefit of the presence of receiver SI at the transmitter side. This question is intimately related to the feedback scenarios in strategic communication: if the receiver has the option of conveying its side information to the transmitter, should it choose to do so? Let us define $D_E^{RSI}$ and $D_D^{RSI}$ as the distortions of the transmitter and the receiver in the setting where SI is also available at the transmitter. The following theorem states that in the Q-G setting, the presence of the receiver SI at the transmitter is not useful to the transmitter or to the receiver.

**Theorem 6.** *In the Q-G setting, the following holds:*

$$D_E^{RSI} = D_E^{SI}, \quad D_D^{RSI} = D_D^{SI}.$$

*Proof.* We begin by showing optimality of linear strategies in these setting where SI is available at both agents. This problem simplifies to one without any SI, analyzed in Theorem 1, since the transmitter can operate on $(X - \mathbb{E}\{X|W\}, \theta - \mathbb{E}\{\theta|W\})$ as the effective $(X, \theta)$ pair and due to jointly Gaussian statistics, $(X - \mathbb{E}\{X|W\}, \theta - \{\theta|W\})$ is statistically independent of $W$. Since the receiver has also access to $W$, the problem is equivalent to minimizing $\mathbb{E}\{(X + \theta - \hat{X})^2\}$ for some $k \in \mathbb{R}$. Following the steps in the proof of Theorem 1, we can show optimality of linear mappings for this modified problem which implies their optimality for the original problem.

Given that $Y = X + a\theta + bW$ for some $a, b \in \mathbb{R}$, and $W$ is also available at the receiver, and the receiver can eliminate $W$ from $Y$ without any change to its estimate (note also that there are no constraints on $Y$ such as a power constraint). Hence, $Y = X + a\theta + bW$ yields distortions ($D_E^{RSI}$ and $D_D^{RSI}$) identical to the ones achieved by $Y = X + a\theta$ ($D_E^{SI}$ and $D_D^{SI}$). □

**Remark 2.** *An essential step in this argument is the absence of constraint associated with $Y$. A constraint on $Y$, such a power constraint, i.e., $\mathbb{E}\{Y^2\} \leq P$ for some $P \in \mathbb{R}^+$ renders the realization of $Y$ useful to the transmitter and also to the receiver as shown in Section V.*

## IV. STRATEGIC WYNER-ZIV PROBLEM

This section focuses on the Wyner-Ziv problem [4] in the strategic settings. The following theorem, whose proof directly follows from standard arguments, states the achievable rate-distortion region ($R, D_E, D_D$ triple), denoted by $\mathcal{RD}_S^{SI}$.

**Theorem 7.** $\mathcal{RD}_S^{SI}$ *is the convex hull of the set of all triplets* $(R, D_E, D_D)$ *for which there exist a function* $h : \mathcal{X} \times \mathcal{W} \to \hat{\mathcal{X}}$ *and a conditional distribution* $p(Y|X, \theta)$ *such that*

$$R \geq I(X, \theta; Y) - I(Y; W) \quad (15)$$
$$D_E \geq \mathbb{E}\{d_E(X, \theta, h(Y, W))\} \quad (16)$$
$$D_D \geq \mathbb{E}\{d_D(X, h(Y, W))\} \quad (17)$$

In general, in non-strategic information theoretic settings, side information has two types of benefits for the receiver, demonstrated in Theorem 7 (for a detailed analysis, see [12, Section 11] and [13]). The first one is *estimation* benefit, which corresponds to the receiver using $W$ (in addition to $Y$) to generate $\hat{X}$, as shown in (16) and (17). This benefit also exists in the single-letter case. The second one, namely the *rate reduction* benefit only exists in the information theoretic setting, and is demonstrated by the term $I(Y; W)$ in (15). In non-strategic settings, the encoder makes $Y$ correlated with $W$ to maximize this rate reduction. However, in strategic settings, there exist problem parameters that render $Y$ independent of $W$ due to differences in $d_E$ and $d_D$, hence make $I(Y; W)$ vanish. This observation plays a pivotal role in the noisy equilibrium with SI setting.

Next, we extend our analysis to the Q-G setting, as shown in Figure 1, where $X, \theta, W$ is jointly Gaussian. The following theorem characterizes the forward test channel that achieves the $\mathcal{RD}_S^{SI}$.

**Lemma 2.** *In the Q-G setting, $\mathcal{RD}_S^{SI}$ is achieved by*

$$Y = X + \beta(R)\theta + S$$

*where $S \sim \mathcal{N}(0, \sigma_S^2)$ is statistically independent of $X$, $\theta$ and $W$. The equilibrium coefficient $\beta(R)$ as well as $\sigma_S^2$ depend on the allowed rate.*

*Proof.* The fact that jointly Gaussian $X, \theta, Y$ achieves $\mathcal{RD}_S^{SI}$ follows from Lemma 1 and the entropy maximization property of jointly Gaussian distribution subject to second order constraints [14]. From the definition of the problem, we have the natural Markov chain $Y - (X, \theta) - W$ (see e.g., [15]). Hence, we have

$$Y = X + \beta\theta + S \quad (18)$$

for some $\beta \in \mathbb{R}$ and $S \sim \mathcal{N}(0, \sigma_S^2)$ is independent of $X, \theta$ and $W$. Plugging (18) into (15), we have (19). and into (16), we obtain (20). noting that $h(Y, W) = \mathbb{E}\{X|Y, W\}$ due to quadratic $d_D$ and is linear due to jointly normal $X, Y, W$. Using (19), we have

$$\frac{\sigma_S^2}{\sigma_X^2} = \frac{1}{2^{2R}-1}\left(1 + \beta^2 r_\theta + 2\beta\rho_{X\theta} - \frac{(\rho_{XW}+\beta\rho_{\theta W})^2}{r_W}\right) \quad (21)$$

We next note that the objective of the encoder is to minimize $D_E$ over the possible choices of $\beta$, which is equivalent to maximizing

$$J(\beta) = -\frac{(1+\beta\rho_{X\theta})(\beta^2 r_\theta + 2\beta\rho_{X\theta})}{1+\beta^2 r_\theta + 2\beta\rho_{X\theta} + \frac{\sigma_S^2}{\sigma_X^2}} \quad (22)$$

over $\beta$. Plugging (21) into (22) we observe that $\beta^* = \arg\max J(\beta)$ depends on $R$. $\square$

**Remark 3.** *In Theorem 3, the compression coefficient $\beta$ is independent of the allowed rate, and identical to the equilibrium coefficient $\alpha$ in Theorems 1 and 5. Here, due to SI, particularly, the $I(Y; W)$ term, $\beta$ depends on the allowed rate, and is obviously different from $\alpha_{SI}$ in Lemma 1 where there is no rate constraint.*

Next, we analyze the benefit of the presence of SI at the transmitter side. A well-known result in networked source coding is the "no rate loss" result of Wyner and Ziv stating that there is no loss of not having access to the receiver SI at the transmitter. At first sight, it might seem that due to the strategic aspect of the problem at hand, the presence of this SI should help to the transmitter (or even to the receiver). The following theorem states that this intuition is not correct, specifically, there is no benefit of the presence of the receiver SI at the transmitter side.

**Theorem 8.** *In the Q-G setting, the following holds:*
$$\mathcal{RD}_S^{SI} = \mathcal{RD}_S^{RSI}.$$

*Proof.* We first note that, following the arguments in the proof of Theorem 6, the transmitter SI does not affect the distortions ($D_E$ and $D_D$). When SI is available at both ends, it can be shown using the arguments in [16] and Theorem 3 that the rate expression simplifies to $R = \min I(X, \theta; Y|W)$ where minimization is over all conditional probability distributions $p(Y|X, \theta, W)$, while when SI is only available at the encoder we have the same minimization over $p(Y|X, \theta)$. Hence, the only difference is due to the additional Markov chain constraint $Y - (X, \theta) - W$, it is well-known that for jointly Gaussian variables this constraint is always satisfied (see e.g., [4], [15]), hence does not affect the minimization. $\square$

## V. NOISY EQUILIBRIUM

In this section, we analyze Q-G noisy equilibirum with the receiver SI. First, we investigate the optimal single-letter strategy within the set of affine strategies. The

**Lemma 3.** *Optimal linear strategies at the noisy Q-G setting with SI are*

$$g(X, \theta) = \sqrt{\frac{P_T}{\sigma_X^2(1+2\alpha_{SI}\rho_{X,\theta}+\alpha_{SI}^2 r_\theta)}}(X + \alpha_{SI}\theta), \quad (23)$$
$$h(Y, W) = \mathbb{E}\{X|Y, W\} \quad (24)$$

*where $\alpha_{SI}$ satisfies the equilibrium in Theorem 1.*

The proof of Lemma ?? follows from standard minimum mean-squared error (MMSE) computations, very similar to

$$R = \frac{1}{2} \log \left( 1 + \frac{\sigma_X^2}{\sigma_S^2} \left( 1 + \beta^2 r_\theta + 2\beta \rho_{X\theta} - \frac{(\rho_{XW} + \beta \rho_{\theta W})^2}{r_W} \right) \right) \tag{19}$$

$$D_E = \sigma_X^2 \left( 1 + 2\rho_{X\theta} + r_\theta - \frac{(1 + \beta \rho_{X\theta})(\beta^2 r_\theta + 2\beta \rho_{X\theta})}{1 + \beta^2 r_\theta + 2\beta \rho_{X\theta} + \frac{\sigma_S^2}{\sigma_X^2}} \right) \tag{20}$$

derivation of $D_E$ in the proof of Lemma 2, and is omitted here. Next, we present our main result pertaining to this setting.

**Theorem 9.** *In strategic, noisy Q-G setting with SI, single-letter linear strategies stated in Lemma 3 are Shannon sense optimal if and only if*

$$\rho_{X,W} = -\rho_{\theta,W} \beta(R), \tag{25}$$

*and*

$$R = \frac{1}{2} \log \left( 1 + \frac{P_T}{\sigma_N^2} \right) \tag{26}$$

*is the capacity of the channel.*

*Proof.* Equating the outer bound obtained by simply applying data processing inequality $R_S^{SI}(D) = C(P)$ to the inner bound that results in the linear mapping in Lemma 3, we get a matching condition which implies that for the Shannon sense optimality, the communication channel in Figure 1 must be identical to the R-D test channel provided in Lemma 2. Note that $\alpha_{SI}$ does not depend on the channel parameters $P_T$ or $\sigma_N^2$. However, $\beta(R)$ depends on the rate, and hence on the channel parameters, due to (26). The only way to make the R-D test channel identical to the actual one is to operate at the rate point where $\beta(R) = \alpha$. From Theorem 7, $\beta(R) = \alpha$ implies that $I(Y; W) = 0$ which is equivalent to statistical independence of $Y$ and $W$. Since all variables are jointly Gaussian with zero mean, the statistical independence implies uncorrelated variables, hence we have (25). □

**Remark 4.** *Theorem 9 does not preclude the possibility of optimality of the mappings in (24) within the set of single-letter strategies even if they do not satisfy (25) in which case they are strictly suboptimal in the Shannon sense (i.e., among $n$-letter strategies).*

## VI. Discussion

In this paper, we have analyzed the impact of side information on strategic compression and source-channel coding problems. Particularly, we have shown that the equilibrium for the quadratic-Gaussian setting with receiver side information admits linear optimal strategies, if there is no channel noise present. Otherwise, i.e., for the noisy case, it does so for the very specific, matched case of the channel noise, the allowed power and the joint statistics of source-private information-side information. Some future directions for research on this general class of problems include a detailed study of vector and networked extensions, and applications of the developed strategic communication framework to other problem areas.